\documentclass[twocolumn]{revtex4}
\usepackage[latin1]{inputenc}
\usepackage{amsmath}
\usepackage{graphicx}
\usepackage{subfigure}
\usepackage{verbatim}%for comments

\newcommand{\strength}{g}

\begin{document}

\title{Doppler-Free Spectroscopy of Weak Transitions:\\An Analytical Model Applied to Formaldehyde}
\author{M. Zeppenfeld, M. Motsch, P.W.H. Pinkse, and G. Rempe}
\affiliation{Max-Planck-Institut f\"ur Quantenoptik,
Hans-Kopfermann-Str. 1, D-85748 Garching, Germany}

\bibliographystyle{unsrt}
\begin{abstract}
Experimental observation of Doppler-free signals for weak
transitions can be greatly facilitated by an estimate for their
expected amplitudes. We derive an analytical model which allows the
Doppler-free amplitude to be estimated for small Doppler-free
signals. Application of this model to formaldehyde allows the
amplitude of experimentally observed Doppler-free signals to be
reproduced to within the experimental error. %
% 39.30.+w  spectroscopic techniques
% 42.62.Fi  Laser spectroscopy
\end{abstract}

\maketitle

\section{Introduction}
The investigation of absorption lines in atomic and molecular gases has been a fundamental tool for the study of the properties of matter. Originally, two different properties of absorption lines could be measured, their frequency and their integrated strength. With the advent of narrow-bandwidth optical light sources, i.e. lasers, the shape of measured absorption lines is no longer necessarily determined by the spectral resolution of available spectrometers, but by the properties of the absorbing gas, thus making a third quantity, the lineshape, accessible to measurement. In many situations the measured lineshape is the Doppler profile, due to the velocity distribution of particles in a thermal gas, which reveals very little about the properties of the absorbing gas. On the other hand, Doppler broadening can be circumvented by making use of nonlinear effects, causing a Lamb dip~\cite{Lamb64,Szoke63,McFarlane63}, or Doppler-free signal, to appear in the spectrum, whose lineshape depends on more than mass and temperature of the absorbing particles. As a result, observation of the Doppler-free signal not only allows the transition frequency to be determined more precisely, but also discloses additional information about the observed particle.

For strong atomic and molecular transitions, observation of the Doppler-free signal is relatively straightforward. Quite the opposite is true for sufficiently weak transitions. For a Doppler-free signal eluding a straightforward observation, an estimate for the amplitude of the Doppler-free signal can be of great value. For this purpose we have developed an analytical model to describe the amplitude of a Doppler-free signal.

The model is presented in section \ref{section theory}. Section \ref{section theory} concludes with a discussion on how to obtain a maximum Doppler-free signal. In section \ref{section experiment} we present the experimental setup which we used to measure the Doppler-free signals of four lines in the $\tilde{X}^1A_1\to\tilde{A}^1A_2$ electronic transition of formaldehyde. Use of formaldehyde is motivated by our research on cold polar molecules~\cite{Rangwala03,Rieger05} where formaldehyde has several advantages: a large linear Stark shift, a high vapor pressure and a relatively accessible electronic transition in the near ultraviolet~\cite{Dieke34}. Formaldehyde is one of the best studied polyatomic molecules~\cite{HerzbergIII,Clouthier83} and is still topic of active research~\cite{Pope05}. Nevertheless, Doppler-free spectroscopy of formaldehyde has not been reported to our knowledge. Finally, in section \ref{section comparison}, we demonstrate the validity of the theoretical model based on the data for formaldehyde.

\begin{comment}
not the strength of the entire line but the amplitude of the Doppler-free signal.
amplitude of Doppler-free signal exists

Setting up a theoretical model to describe the Doppler-free signal is of course useful to investigate the underlying nonlinear mechanisms causing the Doppler-free signal. In addition, for spectroscopy of very weak transitions, where observation of Doppler-free signals is difficult, a theoretical value for the expected amplitude of the Doppler-free signal can be very helpful for the design of an experiment, this being the motivation for the present work.

\end{comment}

\section{Theoretical Model}\label{section theory}
\begin{figure}[tp]
\centering
\includegraphics[width=.5\textwidth]{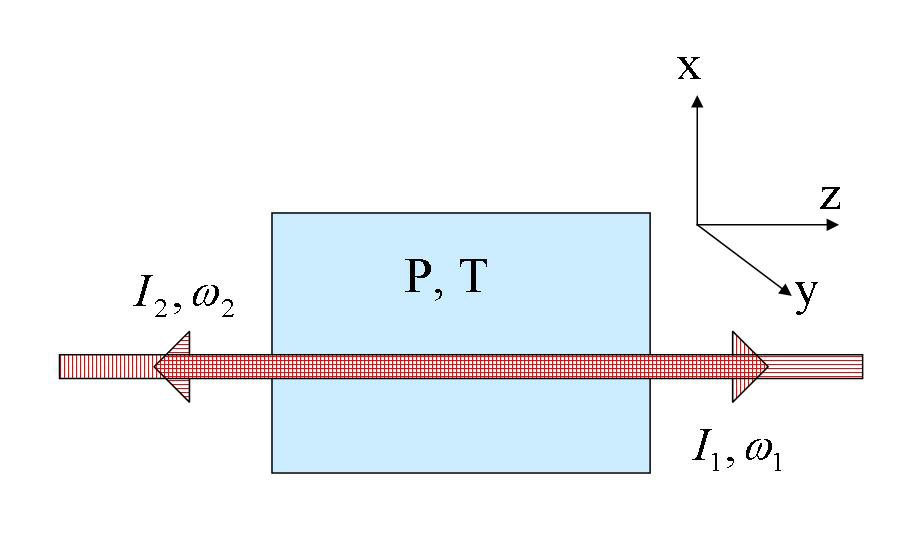}
\caption{Basic setup for Doppler-free spectroscopy. Two counterpropagating laser beams of integrated intensity $I_j$ $(j=1,2)$ and frequency $\omega_j$ pass a cell filled with a gas of pressure $P$ and temperature $T$. In contrast to common practice, we make no distinction between pump and probe beam.}\label{theory setup}
\end{figure}
In this section we present the model to describe the amplitude of the Doppler-free signal which would be obtained for a setup as shown in Fig. \ref{theory setup}. Two monochromatic laser beams of frequencies $\omega_1\approx\omega_2$ with integrated intensities $I_1(z)$ and $I_2(z)$ propagate in the $+\hat{\mathbf{z}}$ and $-\hat{\mathbf{z}}$ direction respectively through a gas at pressure $P$ and temperature $T$. Some atomic or molecular species in the gas has an optical transition of frequency $\omega_0$ in the vicinity of the laser frequencies from some lower state $X$.

The laser beams are assumed to be perfectly overlapping Gaussian beams with a common beam radius of $w(z)$. The time-averaged electric field strength of the laser beams is then given in the paraxial approximation by
\begin{equation}\begin{split}
&|\mathbf{E}_j(\mathbf{x})|=\frac{1}{w(z)}\sqrt{\frac{2I_j(z)Z_0}{\pi}}e^{-\frac{x^2+y^2}{(w(z))^2}}\\
&\mbox{with}\hspace{.5cm}j=1,2,\hspace{.5cm}\mathbf{x}=(x,y,z).
\end{split}\end{equation}
We use $Z_0=\sqrt{\frac{\mu_0}{\epsilon_0}}$, the impedance of free space.

The probability per second for a particle in state $X$ to absorb a photon from laser beam $j$ is given by
\begin{equation}
\frac{\strength}{(\omega_j-\omega_0')^2+(\frac{\Gamma}{2})^2}|\mathbf{E}_j(\mathbf{x})|^2,
\end{equation}
where $\Gamma$ is the homogeneous linewidth of the transition and $\strength$ is a constant which quantifies the transition strength. For a particle moving with $z$-velocity $v_z$, the transition frequency is shifted to $\omega_0'=\omega_0(1\mp\frac{v_z}{c})$ in the lab frame due to the Doppler effect. The top (bottom) sign will always refer to laser beam 1 (2).

If we now assume that $\rho(\mathbf{x},\mathbf{v})$, the phase-space density of molecules in state $X$, is equal to $\rho_0(\mathbf{v})$, the density in thermal equilibrium, we could easily derive the Beer-Lambert law of absorption proportional to laser power. On the other hand, $\rho(\mathbf{x},\mathbf{v})$ is decreased from $\rho_0(\mathbf{v})$ due to optical pumping via excitation by the laser and decay into other states, which is after all the basis for the Doppler-free signal. The steady-state value of $\rho$ is determined by competition of this depopulation with the two processes which cause $\rho$ to reapproach $\rho_0$: state-changing collisions between particles and diffusion of particles in state $X$ into the laser beams from outside the laser beams. The sum of these effects can be modeled by a partial differential equation involving the time rate of change of $\rho$,
\begin{equation}\label{density rate equation}
\frac{\partial\rho}{\partial t}=-\mathbf{v\cdot\nabla}\rho-\frac{1}{\tau_{sc}}(\rho-\rho_0)-c_1e^{-2r^2/(w(z))^2}\rho,
\end{equation}
with
\begin{equation}\begin{split}\label{c_1 and r}
c_1&=\sum_{j=1}^2\frac{\strength}{(\omega_j-\omega_0')^2+(\frac{\Gamma}{2})^2}\,\frac{2I_j(z)Z_0}{\pi(w(z))^2},\\
r&=\sqrt{x^2+y^2}.
\end{split}\end{equation}
The first term in Eq. (\ref{density rate equation}) is due to motion of the particles, the second term is due to state-changing collisions, and the third term is due to depopulation of state $X$ by the laser beams. $\tau_{sc}$ is the average time between state-changing collisions. Note that $\rho_0(\mathbf{v})$ can be expressed in terms of the number of molecules per unit volume $n$, the probability $p_1$ for a molecule to be in state $X$ and an average thermal velocity $\tilde{v}$ as
\begin{equation}\begin{split}\label{rho0}
&\rho_0(\mathbf{v})=\frac{p_1n}{(\pi\tilde{v}^2)^{3/2}}e^{-\frac{|\mathbf{v}|^2}{\tilde{v}^2}},\\
&\mbox{with}\hspace{.5cm}\tilde{v}=\sqrt{\frac{2k_BT}{m}},\hspace{.5cm}n=\frac{P}{k_BT}.
\end{split}\end{equation}
The effect of collisions between particles is highly simplified. In particular, elastic collisions between particles are not taken into account. Consideration of a more complex model is certainly possible\,\cite{Berman75,Bennett80}. On the other hand, use of this simple model allows the steady-state value of $\rho(\mathbf{x},\mathbf{v})$, given by $\frac{\partial\rho}{\partial t}=0$, to be expressed analytically. To do so we define the function $y(t)=\rho(\mathbf{x}+\mathbf{v}t,\mathbf{v})$ and note that this allows us to rewrite Eq. (\ref{density rate equation}) as
\begin{equation}\label{dydt}
\frac{dy}{dt}=-\frac{y-\rho_0}{\tau_{sc}}-c_1e^{-2r^2/w^2}y,
\end{equation}
where $c_1$, $r$ and $w$ are considered functions of $t$ by replacing $\mathbf{x}=(x,y,z)$ by $\mathbf{x}+\mathbf{v}t$.

Eq. (\ref{dydt}) is an inhomogeneous linear first order differential equation of the form $\frac{dy}{dt}+f_1(t)y=f_2(t)$ whose solution can be expressed analytically for arbitrary functions $f_1(t)$ and $f_2(t)$\,\cite{Zauberbuch} and is given in our case, evaluated at $t=0$, by
\begin{equation}\begin{split}\label{rho(x,v)}
\rho(\mathbf{x},\mathbf{v})&=\int_{-\infty}^0\left\{dt'\frac{\rho_0(\mathbf{v})}{\tau_{sc}}e^{t'/\tau_{sc}}\times\right.\\
&\left.\exp\left[-c_1\int_{t'}^0\exp\left(-\frac{2|\mathbf{x}+\mathbf{v}t''|^2}{w(z)^2}\right)dt''\right]\right\}.
\end{split}\end{equation}
Multiplying this density by the probability to absorb a photon and the photon energy and integrating over the laser cross section $\mathbf{x}_t=(x,y)$ as well as over all velocities $\mathbf{v}$ yields the decrease in laser power per distance,
\begin{equation}\label{dIdz}
\left|\frac{dI_j}{dz}\right|=\hspace{6cm}
\end{equation}
$$\int d\mathbf{x}_t\,d\mathbf{v}\,\rho(\mathbf{x},\mathbf{v})\frac{\strength\hbar\omega_0}{(\omega_j-\omega_0')^2+(\frac{\Gamma}{2})^2}\,\frac{2I_jZ_0}{\pi w^2}\,e^{-2r^2/w^2}.$$
We note that we have neglected distortion of the laser beam due to the variation in absorption over the laser beam cross section. Eq. (\ref{dIdz}) together with Eq. (\ref{rho(x,v)}), considered as a function of laser frequency, represents the absorption lineshape including the Doppler-free signal.

With the following additional assumptions, all integrals can be expressed analytically. First and foremost, the relative depopulation of molecules in state $X$ must be small. Primarily, this allows us to linearize the outer exponent in Eq. (\ref{rho(x,v)}) around zero. Additionally, we treat the system as quasi two-dimensional, assuming the beam to be locally invariant under translation in the $z$-direction. This is justified as long as the variation of the beam power and the beam radius in the $z$-direction is small on the scale of the beam radius. Finally, the Doppler-free linewidth must be significantly smaller than the Doppler-broadened linewidth, allowing the Gaussian velocity distribution of molecules to be factored out of integrals. The absorption per distance is then given by
\begin{equation}\label{dIdz result}
\left|\frac{dI_j}{dz}\right|=\alpha\,e^{-\left(\frac{c\Delta\omega_j}{\tilde{v}\omega_0}\right)^2}I_j(z)\times\left.\hspace{3.5cm}\right.\\
\end{equation}
$$\left[1-\frac{\strength Z_0}{4\tilde{v}w}\,{\rm L}\!\left(\frac{w}{\tilde{v}\tau_{sc}}\right)\sum_{j'=1}^2\frac{I_{j'}(z)}{\left(\frac{\Delta\omega_j-\Delta\omega_{j'}}{2}\right)^2+\left(\frac{\Gamma}{2}\right)^2}\right],$$
$$\mbox{with}\hspace{1cm}\alpha=p_1n\frac{2\sqrt{\pi}\strength\hbar
cZ_0}{\tilde{v}\Gamma}\hspace{0.5cm},\hspace{0.5cm}\Delta\omega_j=\pm(\omega_j-\omega_0).$$
The function ${\rm L}(s)$ is the Laplace transform of a Lorentzian function and can be written in terms of Si$(s)$ and Ci$(s)$ which are sine and cosine integrals respectively\,\cite{Zauberbuch},
\begin{equation}\begin{split}
{\rm L}(s)=&\frac{2}{\pi}\int_0^\infty dt\,\frac{e^{-st}}{1+t^2}=\\
&\cos(s)\left(1-\frac{2}{\pi}{\rm Si}(s)\right)+\frac{2}{\pi}\sin(s){\rm Ci}(s).
\end{split}\end{equation}
A basic interpretation of the result, Eq. (\ref{dIdz result}), follows. The term in front of the bracket times the $1$ inside the bracket is Beer Lambert's law: absorption equals the absorption coefficient $\alpha$ times the laser power times a Gaussian spectral profile. The second term in the bracket is the amplitude of the Doppler-free signal, relative to the Doppler-broadened signal. This term is proportional to the laser power and has a Lorentzian spectral profile for $j\ne j'$ as expected. When the assumptions used in the model hold, the second term is always much less than one.

Application of our model to an experiment requires Eq. (\ref{dIdz result}) to be integrated along the laser beam. This integration incorporates two otherwise neglected effects into the model, decrease in beam power along the beam due to absorption and variation in beam diameter due to diffraction. Due to the presence of $w$ as part of the argument of the function ${\rm L}(s)$ in particular, the $z$-integration is no longer solvable analytically, but can easily be implemented numerically.

\subsection{Experimental Implications}\label{implications}
As noted in the introduction, the model we developed is useful in particular as a guide for maximization of a weak Doppler-free signal. For most of the experimentally controllable parameters, the effect on the Doppler-free signal amplitude is relatively obvious. Increasing the laser power and decreasing the laser beam diameter increases the Doppler-free signal whereas increasing the total absorption length $L$ increases the Doppler-free signal at least up to the point where the optical depth $\alpha\,L$ approaches unity. This is easily verified with help of Eq. (\ref{dIdz result}). In particular, the beam radius $w$ enters both as a $w^{-1}$ term and as a linear term in the argument of ${\rm L}(s)$. Since ${\rm L}(s)$ is a monotonically decreasing function in $s$, both occurrences of $w$ lead to the expected behavior. The reader is reminded that this whole discussion only applies in the limit of a small Doppler-free signal.

The optimal pressure for a large Doppler-free signal is less obvious. While increasing the pressure increases the overall absorption signal, it also decreases the scattering time $\tau_{sc}$, thereby increasing the rethermalization rate and decreasing the depopulation of state $X$, accordingly decreasing the Doppler-free signal. The pressure dependence of the Doppler-free signal is therefore nontrivial and warrants a closer look.

\begin{figure}[htp]
\centering
\includegraphics[width=.5\textwidth]{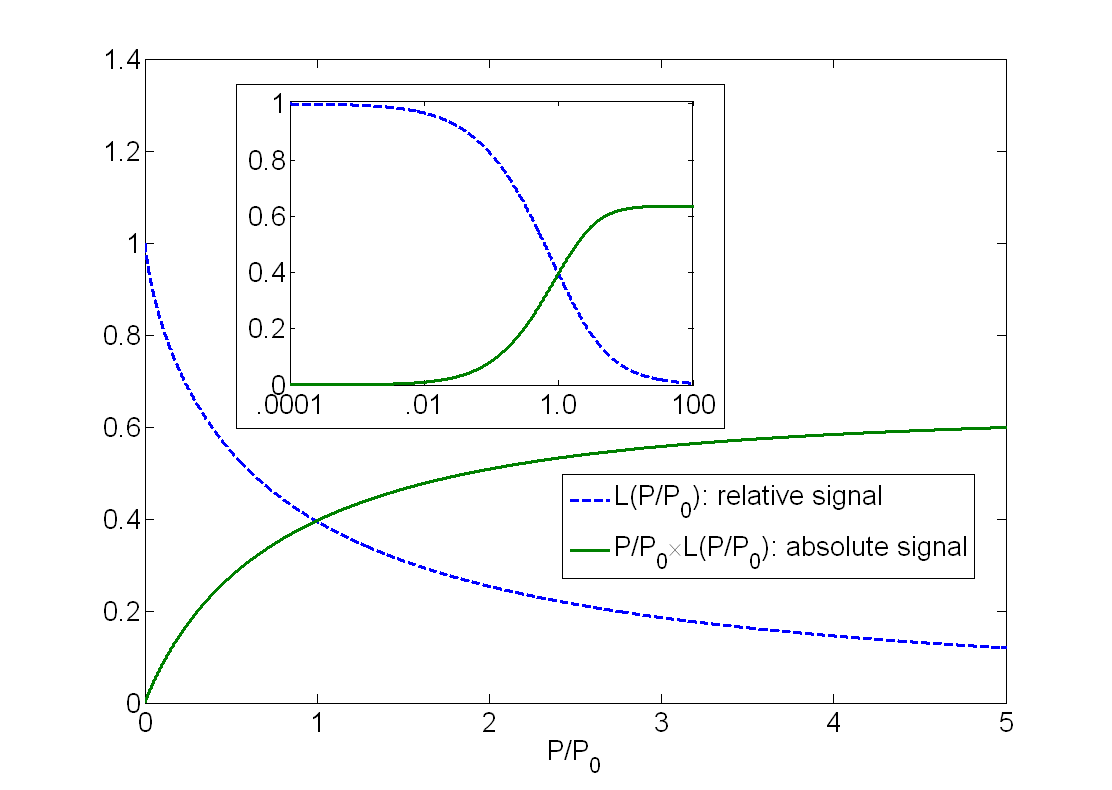}
\caption{Elementary dependence of the absolute Doppler-free signal
amplitude (solid line) as well as the relative amplitude compared to
the Doppler-broadened signal amplitude (dashed line) as a function
of pressure. The inset shows the same data on a logarithmic
scale.}\label{P_L(P)}
\end{figure}
In the present model given by Eq. (\ref{dIdz result}), essentially all parameters other than the fundamental constants show some pressure dependence leading to a nonuniversal pressure dependence. Nonetheless, we can gain significant insight by considering only those parameters which do not approach a finite value in the limit $P\to0$, the molecular density $n$ and the scattering time $\tau_{sc}$, which to first order are proportional and inversely proportional to the pressure respectively. This leads to an absolute Doppler-free signal amplitude of the form $P/P_0\times{\rm L}(P/P_0)$ and an amplitude relative to the Doppler-broadened signal of the form ${\rm L}(P/P_0)$ where $P_0=\tilde{v}\tau_{sc}P/w$ is independent of pressure. These functions are plotted in Fig. \ref{P_L(P)}, and are the basis for the pressure dependence of a Doppler-free signal.

The pressure dependence of the Doppler-free signal in the low-pressure limit $(P\ll P_0)$ and high-pressure limit $(P\gg P_0)$ has a relatively simple explanation. In the low pressure limit, collisions play no role in the degree of depopulation of state $X$. Therefore, the amount of depopulation is pressure independent causing a constant relative Doppler-free signal and an absolute Doppler-free signal linearly increasing as a function of pressure. Conversely, in the high-pressure limit, collisions play the dominant role. Doubling the pressure causes the number of collisions to double and the amount of depopulation to decrease by a factor of two. As a result, the relative Doppler-free signal is inversely proportional to the pressure and the absolute signal approaches a constant. Since a small relative as well as a small absolute Doppler-free signal amplitude hinders observation of the Doppler-free signal, the optimal value for the pressure will generally lie somewhere in the transition region $P\sim P_0$. The pressure $P_0$ is approximately given by the condition that the time of flight through the laser beam is equal to the average time between two collisions.

\section{Experimental Methods}\label{section experiment}
In this section we describe the methods used to measure Doppler-free signals of lines in the $\tilde{X}^1A_1\to\tilde{A}^1A_2$ electronic transition of formaldehyde. Observing Doppler-free lines in this band is difficult for three reasons. First of all, the purely electronic $\tilde{X}^1A_1\to\tilde{A}^1A_2$ transition is a forbidden transition~\cite{Clouthier83}. Only symmetry breaking by means of a change in the vibrational quantum number causes the transition to have a nonzero transition dipole moment. As a result, the transition strength is approximately six orders of magnitude smaller than common values for electronically allowed transitions, hindering depopulation of a lower state and increasing the absorption length by the same amount. Second, the $\tilde{A}^1A_2$ state predissociates causing it to have a lifetime of slightly less than $10\,$ns~\cite{Moore83}. The correspondingly large linewidth further complicates depopulation by diluting available laser power. Since predissociation destroys the formaldehyde molecule, a steady flow of formaldehyde  must be maintained through the spectroscopy chamber during the experiment. Finally the $\tilde{X}^1A_1\to\tilde{A}^1A_2$ is in the UV which limits available laser power.

\begin{figure}[htp]
\centering
\includegraphics[width=.5\textwidth]{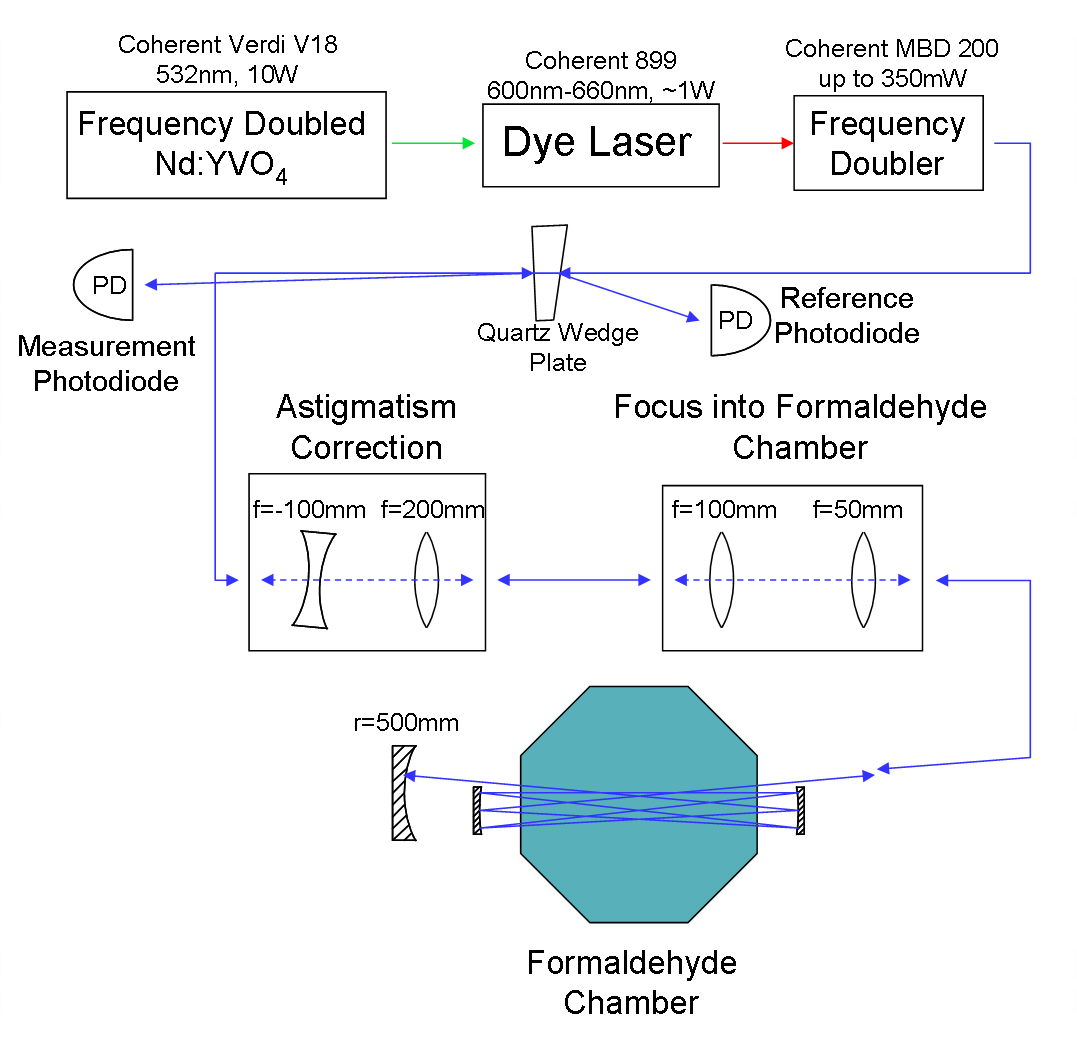}
\caption{Experimental setup used to measure Doppler-free signals of formaldehyde. A frequency-doubled dye laser is focused seven times through a $22.5\,$cm long formaldehyde chamber. The beam is then reflected back on itself and, after retracing its path through the formaldehyde chamber, a part of the beam is diverted by a quartz plate to a photodiode.}\label{laser_setup}
\end{figure}

The setup used to observe Doppler-free lines is shown in Fig. \ref{laser_setup}. A coherent 899 ring dye laser pumped by a frequency doubled Nd:YVO$_4$ laser is frequency doubled to obtain up to $350\,$mW of narrow bandwidth tunable UV light at around $330\,$nm. As pointed out in the last section, the Doppler-free signal can be amplified by increasing the absorption length inside the spectroscopy chamber and by decreasing the laser beam radius. The total absorption length is increased by reflecting the laser beam through the chamber a total of seven times. This leads to a total absorption length of $1.575\,$m. In addition, a set of optics in front of the chamber focuses the beam into the chamber such that the Rayleigh length is approximately equal to the chamber length. Further decreasing the focal size would only decrease the beam diameter at the center of the chamber and would increase the beam diameter everywhere else, leading to an overall decrease in signal.

In order to obtain the optimal beam shape for each pass through the chamber, the mirrors on each side of the chamber are curved so as to match the radius of curvature of the wave fronts. Effectively, the pair of mirrors make up an optical resonator. On the first pass through the chamber, the beam is focused in such a way that it matches the TEM$_{00}$ mode of the effective resonator. The mirrors then automatically refocus the beam to the same optimal profile on each pass through the chamber. While it would certainly be possible to significantly increase the Doppler-free signal by coupling the beam directly into the TEM$_{00}$ mode of the resonator~\cite{Barger69}, this adds the challenge of locking the resonator to the varying laser frequency.

After passing through the chamber seven times, the beam misses the second curved mirror on the opposite side of the spectroscopy chamber and travels to a third curved mirror a bit further on. This mirror is placed at a location such that its radius of curvature again matches the radius of curvature of the beam. By placing the mirror exactly perpendicular to the beam, the laser beam is retroreflected back into itself, having a theoretically perfect overlap with itself. After traveling through the chamber a second time, a part of the beam is extracted using a fused silica glass plate. The glass plate is placed perpendicular to the beam as much as possible to avoid polarization effects.

\begin{figure}[htp]
\centering
\includegraphics[width=.5\textwidth]{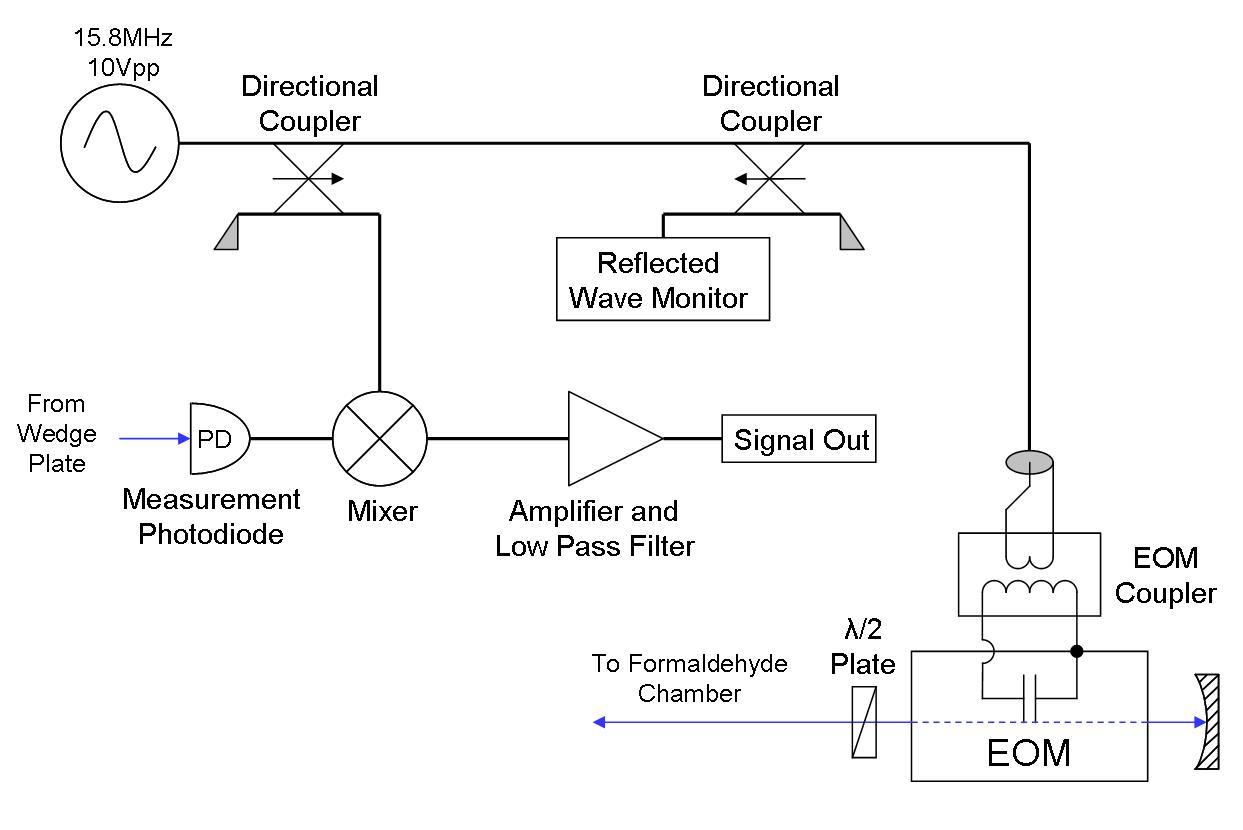}
\caption{Addition to the experimental setup of Fig. \ref{laser_setup} for frequency modulation. An EOM is placed between the formaldehyde chamber and the retroreflecting mirror, frequency modulating the laser beam. The frequency dependent absorption of the gas induces amplitude modulation which is measured by the photodiode.}\label{fm_setup}
\end{figure}

The setup described so far was sufficient to measure Doppler-free signals in the formaldehyde spectrum. However, the obtained signals were small, and could be significantly improved by using frequency modulation techniques\,\cite{Bjorklund80,Bjorklund83}. The necessary changes in the setup are shown in Fig. \ref{fm_setup}. An electro-optical modulator (EOM) used to frequency modulate the laser is placed between the end of the formaldehyde chamber and the retroreflection mirror. The signal from the detector is mixed with the sine wave signal driving the EOM. The DC part of the mixer signal is measured as a function of laser frequency.

\begin{figure}[htp]
\centering
\includegraphics[width=.5\textwidth]{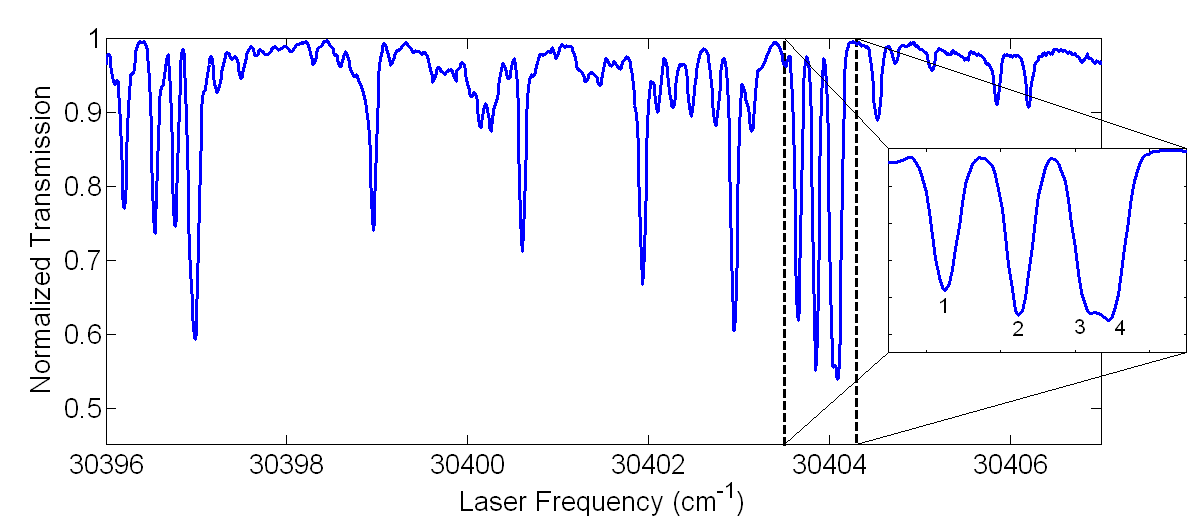}
\caption{Overview of the absorption spectrum of formaldehyde around the lines (1...4) chosen for the Doppler-free study.}\label{spectrum_overview}
\end{figure}

\begin{figure}[htp]
\centering
\includegraphics[width=.5\textwidth]{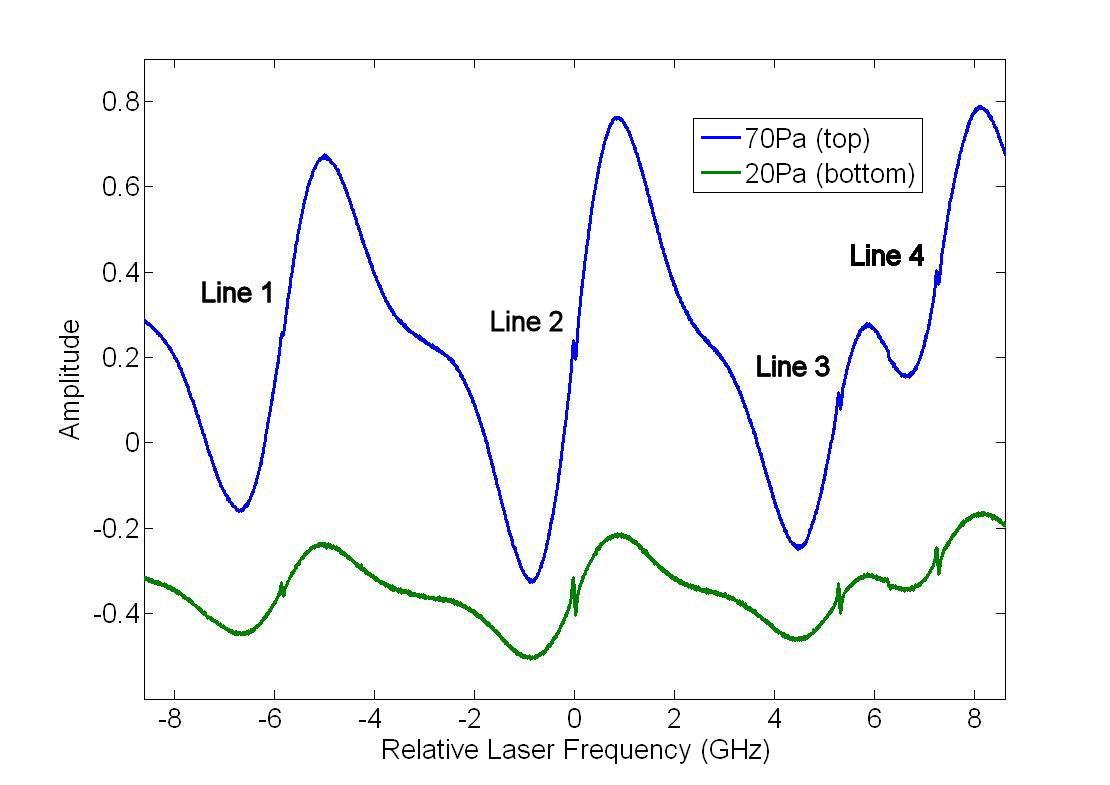}
\caption{FM spectrum of the lines 1...4 of formaldehyde for two
different pressures, including an arbitrary vertical offset for
improved visibility. The sharp Doppler-free signals appear on the
Doppler-broadened lines; all lines appear as dispersive curves due
to the fm technique. }\label{mixer_70Pa_20Pa}
\end{figure}

Fig. \ref{spectrum_overview} shows a small overview of the $2_0^14_0^3$ vibrational band in the electronic transition. The four lines between $30403.5\,$cm$^{-1}$ and $30404.3\,$cm$^{-1}$ were chosen for Doppler-free study since they are the strongest lines in this part of the spectrum. Fig. \ref{mixer_70Pa_20Pa} shows a close-up of the four lines measured using the Doppler-free setup. The Doppler-free signals are clearly visible.

\section{Comparison between Experiment and Theory}\label{section comparison}
In this section we present evidence for the validity of the model for the Doppler-free signal amplitude developed in section \ref{section theory} using the measurements on formaldehyde described in the previous section. The amplitudes of the Doppler-free signals for the four formaldehyde lines were measured at 13 values of the pressure ranging from $1\,$Pa to $100\,$Pa. This corresponds to the transition region between the "low" and "high" pressure regime as described in section \ref{implications}.

\begin{figure*}[ht]
\centering
\subfigure[Line 1]{
\includegraphics[width=.45\textwidth]{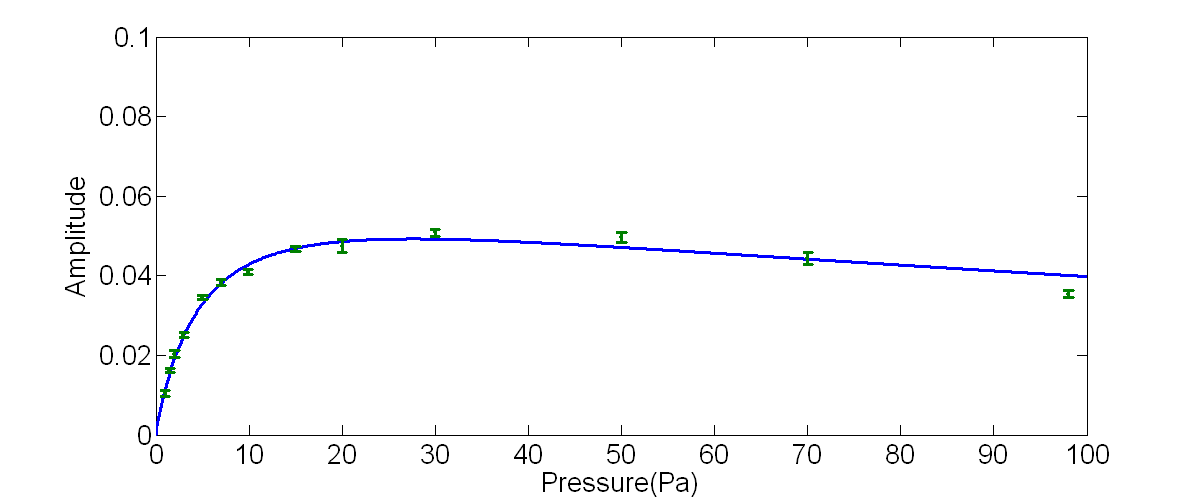}
}
\subfigure[Line 2]{
\includegraphics[width=.45\textwidth]{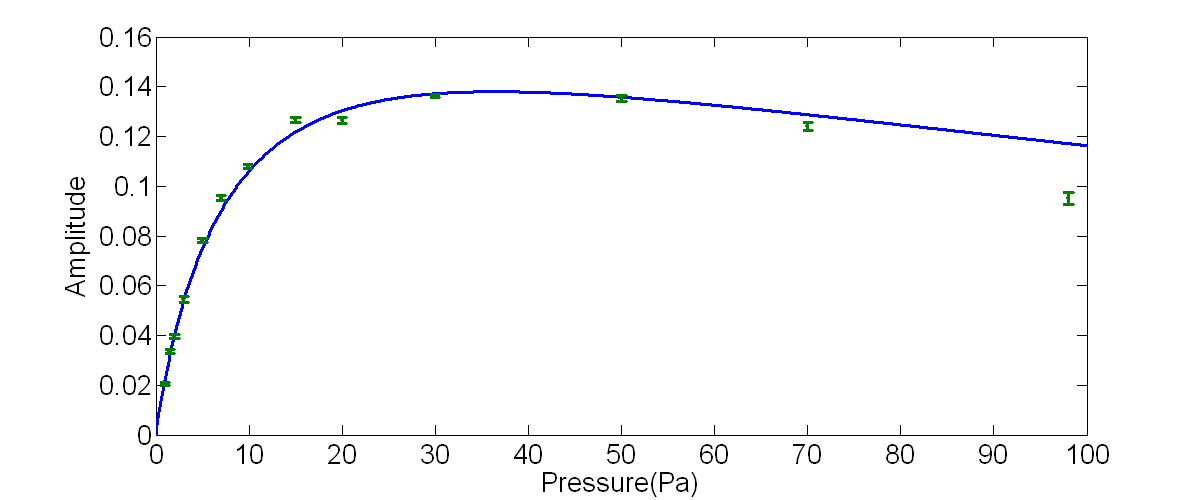}
}
\subfigure[Line 3]{
\includegraphics[width=.45\textwidth]{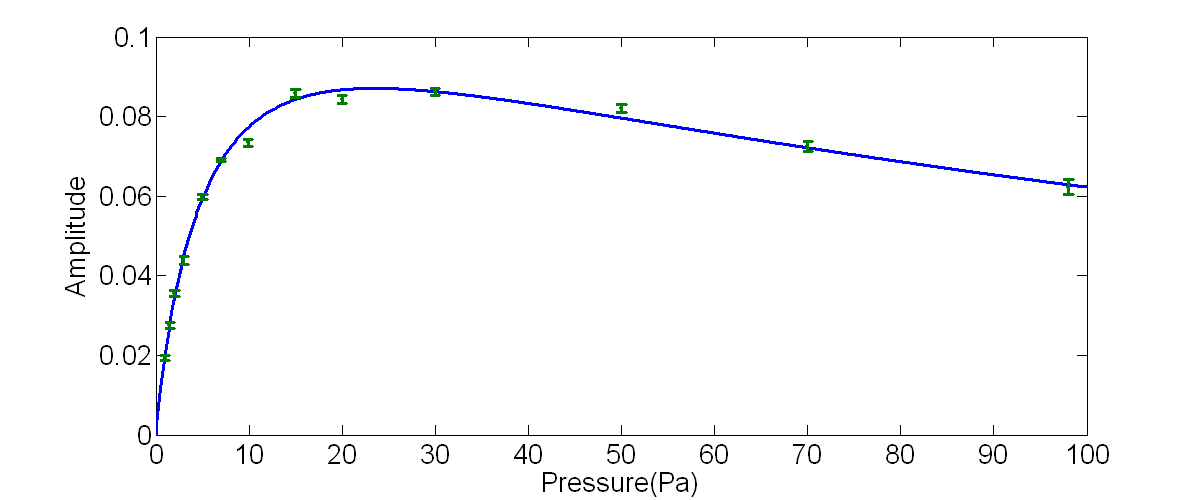}
}
\subfigure[Line 4]{
\includegraphics[width=.45\textwidth]{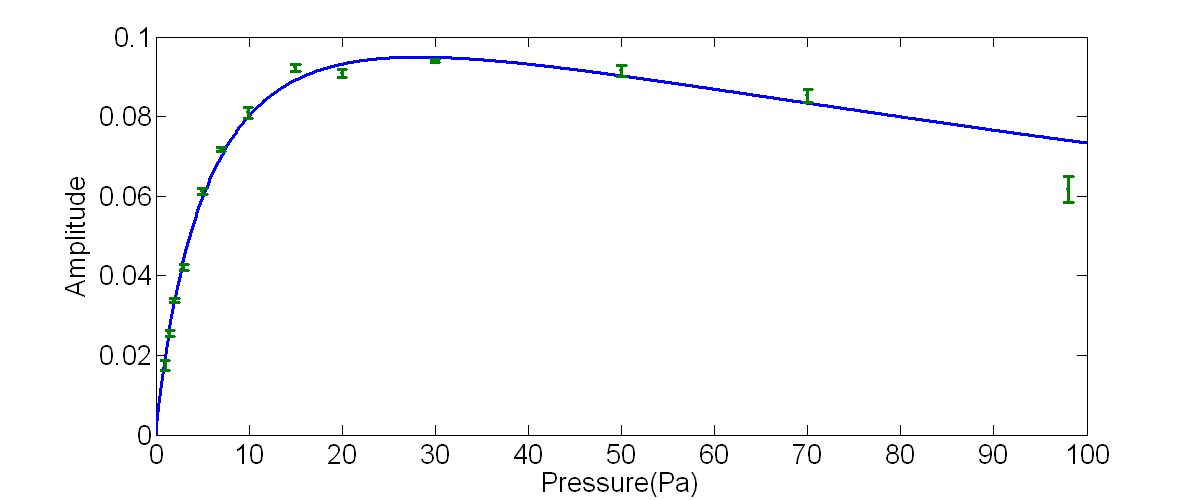}
} \caption{Experimental and theoretical pressure dependence of the Doppler-free signal amplitude for lines 1...4 in (a)...(d), respectively. The experimental data was measured as described in section \ref{section experiment} at a temperature of $294\,$K. The signal decrease at high pressure is due to absorption of the laser beam, causing the average beam power to decrease.}\label{height}
\end{figure*}

\begin{table*}[htb]
\centering
\begin{tabular}{|l|cccc|}
\hline
Lines:&1&2&3&4\\
\hline
Transitions:
&$8_{5,4}\rightarrow9_{6,3}$&$5_{5,1}\rightarrow6_{6,0}$&$7_{5,3}\rightarrow8_{6,2}$&$6_{5,2}\rightarrow7_{6,1}$\\
($J_{K_{-1},K_{+1}}\to J'_{K'_{-1},K'_{+1}}$)
&$8_{5,3}\rightarrow9_{6,4}$&$5_{5,0}\rightarrow6_{6,1}$&$7_{5,2}\rightarrow8_{6,3}$& $6_{5,1}\rightarrow7_{6,2}$\\
\hline
\rule[.1cm]{-.12cm}{.2cm}
Thermal Occupation $p_1$:&$8.74\times10^{-3}$&$7.26\times10^{-3}$&$8.48\times10^{-3}$&$7.99\times10^{-3}$\\
\hline
Linewidth $\Gamma_0$:&$51.0$&$53.3$&$35.5$&$49.3$\\
(MHz)&$\pm0.7$&$\pm0.3$&$\pm0.3$&$\pm0.5$\\
\hline
Pressure Broadening $\Gamma_1$:&$302$&$256$&$287$&$252$\\
(kHz/Pa)&$\pm26$&$\pm10$&$\pm9$&$\pm15$\\
\hline
\rule[.1cm]{-.12cm}{.2cm}
$\alpha_1$ ($10^{-3}$m$^{-1}$\,Pa$^{-1}$):&3.16&3.99&3.44&3.80\\
\hline
\rule[.1cm]{-.12cm}{.2cm}
$\alpha_2$ ($10^{-6}$m$^{-1}$\,Pa$^{-2}$):&5.97&7.32&5.32&4.68\\
\hline
\rule[.1cm]{-.12cm}{.2cm}
Scattering Cross&$5.224\times10^{-18}$&$3.246\times10^{-18}$&$4.794\times10^{-18}$&$4.209\times10^{-18}$\\
Section $\sigma_{sc}$ (m$^2$):&$\pm4.0\times10^{-20}$&$\pm1.2\times10^{-20}$&$\pm2.2\times10^{-20}$&$\pm1.6\times10^{-20}$\\
\hline
\end{tabular}
\caption{Parameters used for the theoretical curves shown in Fig.
\ref{height}. Error bars indicate statistical errors only.
}\label{fit parameters}
\end{table*}

The measured Doppler-free signal amplitude as a function of pressure as well as the theoretical curves for the four lines is shown in Fig. \ref{height}. In obtaining the theoretical curves, two additional effects which were excluded in the discussion in section \ref{implications} were taken into account. First, higher order pressure dependence of the linewidth $\Gamma$ and of the absorption coefficient $\alpha$ were included according to $\Gamma=\Gamma_0+\Gamma_1P$ and $\alpha=\alpha_1P+\alpha_2P^2$. Second, increasing absorption of the laser beam at high pressure resulting in a decreased average laser power in the formaldehyde chamber was included by integrating Eq. (\ref{dIdz result}). This absorption is responsible for the decrease in Doppler-free signal at high pressure seen in Fig. \ref{height}.

Obtaining the theoretical curves requires all parameters in Eq. (\ref{dIdz result}) to be known. These were obtained as follows. The laser beam parameters $I_j(z)$ and $w(z)$ can be acquired from measurements on the laser beam. The laser power is approximately $270\,$mW before entering the spectroscopy chamber. The gas parameters $n$ and $\tilde{v}$ are derived from the temperature and pressure in the spectroscopy chamber according to Eq. (\ref{rho0}).

The occupation probability $p_1$ is equal to the Boltzmann factor
divided by the partition function, taking into account nuclear spin
degeneracy and rotational degeneracy~\cite{HerzbergIII}. To obtain
$p_1$ it is essential to assign the lines, which is non-trivial
because of non-rigid rotor effects and couplings to other states.
The indicated assignment is based on fits to Doppler-broadened
spectroscopic data of the entire $2_0^14_0^3$ vibrational
band\,\cite{Motsch} using genetic
algorithms\,\cite{Meerts:GeneticAlgorithm}. While this fit
determines the $K_{-1}$ quantum number with certainty, the fit
leaves open two possibilities for the order of the four measured
lines. More specifically, exchanging the assignment of lines 1 and 2
and of lines 3 and 4 is not excluded by the line fits. A distinction
between the two orderings is nonetheless possible based on the
progression of frequencies in the $K_{-1}=5\rightarrow K_{-1}=6$
bandhead, leading to the line assignments in table~\ref{fit
parameters}. This order is supported by the fact that the
alternative ordering causes the prediction by our model for the
Doppler-free peak amplitude to fail.

Values for $\Gamma$ and $\alpha$ were obtained for different pressures using fits to curves as presented in Fig. \ref{mixer_70Pa_20Pa} for $\Gamma$ and from fits of absorption as a function of frequency for $\alpha$. The transition line parameters $\Gamma_0$, $\Gamma_1$, $\alpha_1$, and $\alpha_2$ are obtained from the pressure dependence of $\Gamma$ and $\alpha$.

The only parameter which cannot be derived from measurements is $\tau_{sc}$, which is taken as a fit parameter according to $\tau_{sc}=(\sigma_{sc}\tilde{v}n)^{-1}$. The constant $\sigma_{sc}$ is independent of pressure and can be viewed as a scattering cross section. This leads to the excellent agreement between theory and experiment shown in Fig. \ref{height}. The parameters used for the theoretical curves are listed in table~\ref{fit parameters}.

Finally, we estimate the overall accuracy of the experimental
parameters entering the theory. First, we estimate the error in the
pressure reading to be $30$\%, causing the molecule density in the
theoretical model to be inaccurate by the same amount. Second, mass
spectrometer data suggests that the formaldehyde, which dissociates
under UV radiation, is not completely pure, causing the theoretical
values for the Doppler-free amplitude to be underestimated by
approximately $10$\%. Finally, during data acquisition, the laser
power entering the cell was measured with an accuracy of only
$10-20$\%. The beam is further attenuated by $1-2$\% for each pass
through the windows of the spectroscopy chamber, a value which
deteriorates over time as formaldehyde polymerizes on the windows.
Worst case, these errors add up to a factor of two.

In the light of these inaccuracies, the agreement between the theoretical curves and the experimental data in Fig. \ref{height} is perfect. The absolute accuracy of our theoretical model is therefore bounded by the magnitude of the sum of the systematic effects. We have therefore demonstrated that our theoretical model reproduced the absolute amplitude of the Doppler-free peak to within at least a factor of two.

\section{Conclusion}
In summary, we have presented an analytical model to calculate the amplitude of Doppler-free signals. In particular this model gives a relatively simple expression for the Doppler-free signal amplitude. A characteristic pressure dependence of the amplitude of the Doppler-free signal is found. The model is verified by measurements on formaldehyde.

We acknowledge financial support by the Deutsche
Forschungsgemeinschaft (SPP 1116 and cluster of excellence Munich
Centre for Advanced Photonics).

%\bibliography{../MZbib}

\bibliographystyle{unsrt}

\end{document}